\documentstyle[12pt]{article}
\normalsize

\input{psbox.tex}

\def\d{\delta}
\def\e{\epsilon}

\def\j{\psi}

\def\l{\lambda}
\def\m{\mu}
\def\n{\nu}

\def\p{\pi}
\def\q{\theta}

\def\cb{{\cal B}}

\def\pa{\partial}

\def\Hat#1{\rlap{\kern.10em$\widehat{\phantom G}$}#1}
\def\HAt#1{\rlap{\kern.05em$\widehat{\phantom G}$}#1}

\def\cap#1{\rlap{\kern.1em$\widehat{\phantom{G\vrule height.8em}}$}#1{}}
\def\Cap#1{\rlap{\kern.05em$\widehat{\phantom{G\vrule height.8em}}$}#1{}}

\newcounter{sxn}

\newcounter{axn}

\def\br{\begin{thebibliography}{abc}}
\def\er{\end{thebibliography}}

\date{}

\begin{document}
\bibliographystyle{unsrt}
\footskip 1.0cm
\thispagestyle{empty}
\setcounter{page}{0}
\begin{flushright}
SU-4240-534\\
March 1993\\
\end{flushright}
\vspace{10mm}
\centerline {\LARGE SAHA AND THE DYON\footnote {To be published in the
Commemoration Volume celebrating the Birth Centenary of Professor M.N. Saha.}}
\vspace*{10mm}
\centerline {\large A.P. Balachandran}
\vspace*{5mm}
\centerline{\it Department of Physics, Syracuse University,}
\centerline{\it Syracuse, NY 13244-1130}
\vspace*{3cm}
%\baselinestretch{2.0}
\normalsize
\centerline {\large \bf Abstract}
\vglue 6mm
\indent Meghnad Saha occupies a special role in the history of Indian science,
having
been a pioneer in its organization already from the oppressive colonial period
and having left important legacies to post-colonial India like the Saha
Institute of Nuclear Physics.  He is famous for his research in astrophysics,
and has also made important, but less well-known contributions to magnetic
monopole theory.  This article attempts an elementary survey of this theory and
its implications with special emphasis on Saha's work and argues that the
latter has had wide ranging consequences for fundamental physics.
\newpage
\baselineskip=24pt
\setcounter{page}{1}
%\sxn{INTRODUCTION}
\centerline{\large 1}
\vglue 6mm
\indent The interactions of electrically charged matter and the electromagnetic
 field
are described by the Maxwell-Lorentz theory which has the Lorentz force and
Maxwell equations as its principle dynamical equations.  This theory is
immensely successful in describing a wide range of phenomena already in its
classical form.  When it is appropriately first and second quantized, it
becomes capable of accounting for observations all the way from chemistry and
condensed matter physics to Lamb shift and electron-positron $(e^--e^+)$ pair
production.  The typical energies involved at one end of this spectrum are a
few electron volts whereas it costs a minimum of 1 MeV to pair produce
$e^--e^+$.  It is thus a theory which works well for at least six decades in
energy.  Its range of validity is in fact greater than is indicated by these
examples.  It is also a theory accurate to an extent never before encountered
in science.  The agreement between theory and experiment for the anomaly in the
electron gyromagnetic ratio is about one part in $10^7$. Quantum
electrodynamics is one of the best scientific models about
nature ever constructed by humans.
\vglue 8mm
\centerline {\large 2}
\vglue 6mm
\indent It was apparently felt already in the last century by Poincar\'{e},$^1$
Thomson$^2$ and perhaps others as well that there was room for improvement of
electrodynamics despite its empirical success.  Such a speculation becomes
understandable when it is realized that the equations of electrodynamics
contain an intrinsic asymmetry between electricity and magnetism, allowing as
they do for electric monopoles, but not for magnetic monopoles.  It can
therefore be proposed with good reason that just as there are electrically
charged particles which act as sources for  the electric field $\vec E$, so too
must there exist magnetically charged particles which are sources for the
magnetic induction $\vec B$.  In the presence of a particle of charge $e$ at
point $\vec z$, Gauss's law is $\vec \nabla \cdot \vec E(\vec x) =
4\p e \d^3(x-z)$.  In the same way, the equation for $\vec \nabla
\cdot \vec B$ will read $\vec \nabla \cdot \vec B (x)=
4\p g\d^3(x-\vec z)$ when there is a
particle of magnetic charge $g$ at $\vec z$.  The
relativistically covariant version of this equation is
$\pa^{\l}~^{*}F_{\l\n}=4\p j_{\n}^{M}$ where $^*F_{\l\n}$ is the dual of the
Maxwell tensor field $F_{\l\n}$, $j_{\n}^{M}$ is the conserved magnetic four
vector current and we use units where the speed of light $c$ is 1.
In the  work of Poincar\'{e},$^1$ the
dynamics of an electrically and a magnetically charged particle in interaction
were also studied in the  nonrelativistic approximation. There were no doubt
early speculations as well about the physical implications of the existence of
a magnetic monopole.
\vglue 8mm
\centerline {\large 3}
\vglue 6mm
\indent There is one striking consequence of the existence of a magnetic
monopole
already for classical electromagnetic theory.  It is that with a magnetic
monopole
around, it is impossible to introduce a smooth vector potential $\vec A$ for
$\vec B$.  This is because $\vec B=  \vec \nabla\times \vec A$ implies that the
net flux of $\vec B$ over any closed surface is zero by Stokes's
theorem whereas
it must be $4\p g$ for a sphere enclosing magnetic charge $g$.  We may now
remember that the existence of a vector potential $A_\m~ (\m=0,1,2,3)$ is used
in a significant manner for writing a Lagrangian for conventional
electrodynamics.  An electrodynamic theory which has magnetic charges as well
does not therefore admit a simple Lagrangian description. [But see however
ref.3.]

The implications of this remark are not serious for classical physics.  It is
possible to do a great deal of the latter without ever talking about
Lagrangians and Hamiltonians.  Lagrangian and Hamiltonian formalisms are
largely superstructures on the edifice of classical theory.  Perhaps the most
serious exception to the decorative role they play occurs in  classical
statistical mechanics which uses the phase space in a basic manner.  But
electric and magnetic monopoles in interaction do admit a Hamiltonian
description so that the lack of a decent Lagrangian of a conventional sort for
this system is not a serious matter for its classical physics.

But that however is far from being the case in quantum theory, a profound fact
discovered by Dirac in a paper published in 1931.$^{4}$  One way to
sketch Dirac's reasoning about the electric charge-magnetic monopole (or
charge-monopole) system is as follows.  Since a Coulomb magnetic induction does
not admit a vector potential, let us modify it and instead consider
\begin{eqnarray}
\vec {\cb} &=& \vec B_C + \vec B_S\,,\nonumber\\
\vec B_C &=& g \frac{\vec x}{r^3},\,\, r=|\vec x|\,,\nonumber\\
\vec B_S &=& - \vec n~ 4 \p g~ \q(x_3)\d(x_1)\d(x_2),\,\, \vec n = (0,0,1),
\end{eqnarray}
\noindent $\q$ being the standard step function.
In this expression, the Coulomb field $\vec B_C$ of a magnetic monopole (or
monopole for short) at the origin has been augmented by a field $\vec B_S$
concentrated on the nonnegative third axis $x_3 \geq 0,~~x_1=x_2=0$.  This new
$\vec \cb$ is the field of an infinitely thin solenoid on the nonnegative third
axis and is divergenceless.  It can thus be written as $\vec
\nabla \times \vec A$. The dynamics of a charge
in this field can also be described using a Lagrangian.

The field $\vec {\cb}$ is identical to the Coulomb field $\vec {B_C}$ away
from the third axis.  But that is not the case on the
third axis so that without further qualifications or physical inputs, it can
not be regarded as the field of a monopole, but rather must be thought of as
the field of a solenoid.

A brilliant contribution of Dirac to the charge-monopole theory was to
establish that $\vec B_{S}$ is entirely unobservable if the quantization
condition
\begin{equation}
eg = n \frac{\hbar}{2},~n=0, \pm 1,\pm 2,...\,
\end{equation}
\noindent is fulfilled.  The location $L$ of the solenoid has been chosen
to be along the
nonnegative third axis in (1).  But it is obvious that a vector potential
$\vec A$ exists even if this location is along any line from origin to
infinity.  The line of location $L$ of the solenoid is known as the Dirac
string.  The result of Dirac is that the Dirac string can not be observed if
the quantization condition (2) is fulfilled.

This result can be established by semiclassical reasoning as follows.  Let us
for definiteness assume that $L$ is along the third axis as in (1) and imagine
an Aharonov-Bohm experiment$^5$ with the flux on $L$ serving as the flux line.
In this experiment, a coherent beam of particles of charge $e$ and mass $m$
(regarded for simplicity as spinless) are injected through a slit and detected
at a point $\vec x_0$ on a screen as in Fig. 1.
\begin{figure}[hbt]
\begin{center}
\mbox{\psboxto(\hsize;0cm){dirac.eps}}
\end{center}
\end{figure}
The wave function $\j$ at $\vec x_0$ is the sum of two terms in the
semiclassical approximation:
\begin{equation}
\j(\vec x_0) =~ {\rm Overall~ factor}~ \times \left [
e^{iS_{1}/\hbar}+e^{iS_{2}/\hbar}\right ]\,.
\end{equation}

\noindent If $P_i$ are the classical paths from slit to $\vec x_0$, one
behind and the
other before the Dirac string as in Fig. 1, then $S_i$ is the classical action
for path $P_i:$
\begin{equation}
S_i=\int_{P_{i}} \left \{ \frac {1}{2} m \left (\frac {dx_{i}(t)}{dt}\right )^2
- V(\vec x (t) ) + eA_i(\vec x (t)) \frac {dx_{i}(t)}{dt}\right \}\,.
\end{equation}
 \noindent Here we have included a possible potential term (the integral of
$-V$) in the
action which may be required to realise the classical solutions with paths
$P_i$. [Instead of a potential, we can also use  mirrors (or any other suitable
device) to reflect the split beams from the slit so as to realise a path $P_1$
behind and a path $P_2$ before $L$.] The field on the string does not affect
the classical trajectories.  It can therefore be detected only  by its
contribution to the phase difference
of the terms in (3).  This contribution is contained in
\begin{equation}
\exp i \left \{ \frac {e}{\hbar}  \left ( \int_{P_{1}}~-~\int_{P_{2}} \right )
A_i
(\vec x(t)) \frac {dx_{i}(t)}{dt} \right \}
\end{equation}
\noindent
which by Stokes's theorem is
\begin{equation}
\exp i\left \{ \frac {e}{\hbar} \times ~{\rm  Magnetic~ flux ~through ~loop~
\em {C}}~
\right \}
\end{equation}
\noindent where $C$ circles around $L$ as shown in Fig. 1.  Besides
the monopole field
contribution, this flux contains also the string contribution $-4 \p g$, the
shift in the phase difference caused by the flux supported on $L$ being

\begin{equation}
\exp i \left \{ - 4 \p eg /\hbar \right \} \, .
\end{equation}
\noindent If $- 4 \p eg /\hbar$ is $2\p \times$ an integer $n$, then (7) is 1
  and the Dirac
string can not be detected, the answer for the phase difference being the same
whether or not there is the flux supported on $L$.  In other words, according
to Dirac, an electric charge coupled to a potential $\vec A$ with $\vec {\cb}$
as
its curl leads to a  theory of charges and monopoles (and not of charges and
solenoids) provided.

\begin{equation}
eg = \frac {n\hbar}{2}, \,n~ \e ~{\rm the~ set}~ \bf Z ~{\rm of~ integers}~.
\end{equation}

\noindent It is thus the case that we have the quantization rule (8) for
electric and
magnetic charges.

A very significant implication of (8) is that electric charges must be
quantized if there are magnetic monopoles.  Thus suppose that the least nonzero
value of $|g|$ occurs when $g=g_0$.  Then all electric charges are integral
multiples of $\hbar/(2g_0)$ by (8).  Magnetic charges too must come in
quantized
units for a similar reason.  Since experiments indicate that electric charges
are quantized, this prediction of (8) was for a long time considered as a
strong argument in favour of the existence of magnetic monopoles.
\vglue 8mm
\centerline {\large 4}
\vglue 6mm
\indent There is an interesting alternative approach to the derivation of
(8) which is
due to Saha.$^6$   Let us consider the motion of a particle of electric charge
$e$ in the Coulomb field of a particle of magnetic charge $g$.  For
nonrelativistic kinematics, if $\vec z$ is the relative coordinate and $\m$ is
the reduced mass, the Lorentz force equation is

\begin{equation}
\m  \ddot {\vec{z}} = eg~ \frac {\vec{z}\times\dot{\vec{z}}}{|\vec
z|^{3}}\,,
\end{equation}
\noindent the dots denoting time derivatives.  The usual angular
momentum $\m~\vec z \times \dot{\vec{z}}$ is not a constant of motion for (9).
  Instead, as
is well known,
\begin{equation}
\vec J= \m ~\vec z \times \dot{\vec{z}} + eg~ \hat{z},~\hat{z}:=\vec z/|\vec
z|
\end{equation}
\noindent is a constant of motion for (9) and should be regarded as the
angular momentum of
the charge-monopole composite.  There is thus a new term in the angular
momentum pointing along the line joining the charge and the monopole~:
\begin{equation}
\vec J \cdot \hat z = eg \,.
\end{equation}
\noindent The familiar quantization condition for components of angular
momentum is now seen to require the condition (8).

Saha also emphasized that the term $eg ~\hat z$ can be obtained from the
integral
\begin{equation}
\frac {1}{4 \p} \int \left \{\vec x \times \left [ \vec E (\vec x) \times \vec
B (\vec x)\right ] \right \} d^3x
\end{equation}
\noindent which is the contribution of the electromagnetic field to angular
momentum.  This result is originally due to Thomson$^2$.  It can be derived by
substituting the Coulomb fields of a charge and a monopole for $\vec E$ and
$\vec B$ in (12) and performing the integral.

The derivation of the quantization condition (8) from
the properties of angular momentum in quantum theory reveals certain notable
properties of the dyon or the Saha dyon as the charge-monopole composite is
often called.  We can explain them, and bring out their novelty as well, by
first considering a system of two electrically charged particles with no
magnetic charge.  Let us assume furthermore that the particles are spinless.
In that case, the angular momentum or spins of the composite are integral.  In
other words, we can not form a composite of spin 2S with 2S an odd integer (or
spin ``half-integral'') out of spinless constituents.   This remark is correct
for most familiar interactions between particles and also if both particles
have integral or half-integral spin.  There is an obvious modification of this
observation if one of the particles is of integer spin and the second of
half-integer spin.

The story is strikingly different for a Saha dyon.  For a dyon, if $eg$ is an
odd multiple of $\hbar/2$, then the component of angular momentum in the $\hat
z$ direction, and hence the spin of the dyon, is half-integral.  This is so
even though both constituents are spinless (or for that matter, of any integral
or half-integral spin).  It is thus possible to contemplate half-integral spin
or ``spinorial'' composites made up of integral spin or ``tensorial''
constituents.  Saha's derivation of (8) in this manner also emphasizes a very
interesting physical principle.

It is not a priori obvious that such spinorial composites also obey Fermi-Dirac
statistics and hence are fermions.  As dyons have several unusual properties,
one could suspect that their statistics too is unusual.  They may for example
violate the spin-statistics theorem or obey a relatively unfamiliar
parastatistics.  It is known however that the spin-statistics correlation for
dyons is normal and that tensorial (spinorial) dyons obey Bose-Einstein
(Fermi-Dirac) statistics and hence are bosons (fermions).$^7$

Dirac himself\,$^4$ and Tamm soon after him,$^8$ had studied the
Schr\"{o}dinger
equation of the charge-monopole system with spinless constituents in polar
coordinates and thereby found half-integral angular momenta for suitable $eg$
for this composite.  The merit of Saha's approach in contrast to that of these
authors is that it is very simple and reveals the result in an easy manner.
It has since been generalized and is frequently used in research in elementary
particle theory.

Summing up, the work of the above authors and later developments have revealed
a very important physical possibility :  It is conceivable to have spinorial,
fermionic composites made up of tensorial, bosonic constituents.  Later on, we
will have more to say on this remarkable fact originating in dyon theory.
\vglue 8mm
\centerline {\large 5}
\vglue 6mm

In elementary particle physics, the so-called standard model nicely accounts
for properties of strong, electromagnetic and weak interactions.  It is very
successful for energies up to 1 TeV, and may in fact be  good for energies all
the way up to about $10^{15}$ or $10^{16}$ GeV.

The standard model however is unsatisfactory in at least one respect :
 whereas it
successfully unifies electromagnetic and weak interactions, explaining their
origin from a common source, it does not unify all the preceding three
interactions in a similar way.  Many particle physicists therefore suspect that
it will break down at $10^{15}-10^{16}$ GeV, the unity of the three
interactions becoming manifest at such energies.  [The reasons for arriving at
the ``grand unification'' scale of $10^{15}-10^{16}$ GeV are rather technical.]
According to these physicists, there must exist a unified theory of  all these
interactions superseding the standard model which should be used for a
satisfactory description of nature at these energies.  There are at present
several proposed models attempting this unification and collectively known as
Grand Unified Theories (GUT's) or Supersymmetric Grand Unified Theories
(SUSYGUT's).  There is no single GUT or SUSYGUT which we can now say is
closer to
reality than others, but a common expectation is that a unique theory will
eventually emerge as the winner from among the competing GUT's and SUSYGUT's.

It is not necessary to distinguish GUT's from SUSYGUT's for the purposes of
this article.  We will therefore refer only to GUT's.

One feature of GUT's is of particular interest for monopole theory.  It is that
GUT's generically predict the existence of monopoles (and dyons). [For reviews
of GUT monopoles and dyons, see for example ref. 9.]    The prediction of
monopoles by field theories which share the pertinent features of GUT's was
first pointed out by 't Hooft and Polyakov.$^9$  There are differences in
detail between GUT monopoles and those of Dirac, but
there are enough points of similarity that a particle physicist does not always
bother to distinguish these monopoles.  We will accept this practice here.

Monopoles predicted by GUT's are extremely massive, with masses of the order of
the grand unification scale.  They are impossible to create in accelerators.
They could however have been produced during cosmic history.  In that case,
they may be detectable by their flux on earth from outer space, or by
astrophysical observations which look for their effects on magnetic fields in
galaxies or neutron stars.  Experiments along these lines have been done and
are in progress, but they have yet to detect a monopole.  The discovery of a
monopole will be of great significance for fundamental theory.

GUT~ monopoles have many exotic properties.  Just as the Saha dyon, they can be
spinorial fermions, even though their constituent fields are of integer spin.
They also have an astonishing ability to catalyze fast baryon decays,$^{10}$ a
feature which they share with certain GUT ``cosmic strings'' and which will
now be briefly discussed.

GUT's~ generically contain interactions which lead to nonconservation of baryon
(and lepton) number.  In the absence of monopoles, these interactions predict
the decay of protons into leptons plus other particles like mesons and photons.
The life-time for proton decay is typically predicted to be $10^{32}$ years or
longer.  Experiments have not yet detected proton decay.  Its observation will
have as much an impact as the discovery of a monopole on fundamental theory.

Monopoles in GUT's are made of fields which can mediate baryon decays.  It was
discovered in the '80's that as a consequence, a baryon scattering off a
monopole can turn into a lepton plus possible debris like mesons, leaving  the
monopole intact.  The existence of such catalytic processes in itself is not a
surprise.  What is a surprise is that their cross-sections are comparable to
strong interaction cross-sections because of certain powerful attractive
interactions caused by monopole topology.  The upshot is that proton decay
catalyzed by monopoles has a life time of only about $10^{-20}$ seconds (give
or take a few  orders of magnitude) and not $10^{32}$years.  Thus life as we
know it will certainly not exist in an ambience with an abundance of GUT
monopoles.

Monopoles also have implications for the ``standard model'' of the history of
the universe.  It is based on the ``big bang'' alleged to be the beginning of
all history.  In its original version, it predicted the presence of too many
monopoles in outer space leading to contradictions with observations.  This was
among the reasons leading to revisions of this version and to proposals of the
``inflationary'' scenario and its variants.  The interested reader  should
consult a suitable semipopular or technical review of cosmology for further
details.

We can conclude this section as follows:  whereas for electrodynamics, a
monopole was a postulate grafted on to existing theory, that is not the case
for GUT's which generically predict them as stable states.  Their experimental
discovery will therefore have far reaching consequences.

\vglue 8mm
\centerline {\large 6}
\vglue 6mm

The possibility of creating spinorial composites out of tensorial constituents
is a very basic physical result.  It could have been adequately understood by
any physicist after the work of Dirac$^{4}$ and Saha$^6$.  But that is not what
happened, and it was only in the late '60's and the '70's that this result
began to be widely known in the theoretical physics community.  There was at
that time a pervasive feeling too that it hinged on rare topological features
and was not likely to find too many applications.

But all that is changed today especially because of the discoveries of Skyrme
beginning from the late '50's.  [See for example ref. 3 or 11 for a review of
Skyrme's work and related developments.]  He showed in his studies that spin
half excitations can exist in a theory of pions.  He thus proved that spinorial
states can emerge from a tensorial field theory (a variant of the result for
the Saha dyon) and went on to propose that nucleons are twisted topological
lumps in pion fields.  Skyrme's work was revived in the '80's
by Pak and Tze and by the Syracuse group and found universal acceptance after
fundamental contributions by Witten. This model of the nucleon is
called the Skyrmion.

As the Skyrmion is widely known these days at least by name, one supposes that
many physicists also appreciate now that Lagrangians built from tensorial
variables can nevertheless possess spinorial states.

``Geometric Quantization''$^{12}$ is a particular approach to quantization of
classical theories based on powerful mathematical and in particular topological
ideas.  The relatively few physicists actively pursuing this approach have long
appreciated that the Dirac-Tamm-Saha ideas are hardly unique to the
charge-monopole system.  They had in fact found several exceptionally
elementary systems many years ago which we can regard as illustrating these
ideas.  One such example will be mentioned here. [Refs. 3 and 12 can be
consulted for details and more
citations.]  Suppose that one wishes to describe a particle of {\em fixed}\,
spin $j$ by a Lagrangian.  There is then such a Lagrangian based on elements
of the
rotation group SO(3) or on oriented orthonormal frames which are both tensorial
variables.  This is so for $2j$ even {\em or}\, $2j$ odd, the states of course
being spinorial in the latter case.  Particle of fixed half-integral spin thus
illustrate how Lagrangians with tensorial variables can nevertheless have
spinorial states.

There are also many examples of this sort in molecular physics in the
Born-Oppenheimer approximation$^{13}$ and in the collective model approach to
nuclei.$^{14}$   In the former approximation for example, a  molecule can be
described by a Lagrangian based on SO(3).  This description is not unlike the
one commonly used for a rigid rotor.  Now it is easy to demonstrate that the
latter admits quantization with spinorial states [or spinorial quantization].
[Cf. ref. 3, Chapter 13.4.]  For related reasons, the Born-Oppenheimer
Lagrangian too admits spinorial quantization for a great many molecules,
$^{15}$
the spinorial states admitting an interpretation as well in terms of intrinsic
spins of nuclei and electrons.  Molecular and nuclear physicists routinely use
such quantization, but apparently without emphasizing the remarkable
topological and conceptual foundations of quantum physics which allow this
possibility.  If they had paid careful attention to these features, they would
have enriched and significantly extended the Dirac-Tamm-Saha studies and
contributed to the foundations of important research in elementary particle
theory such as that on the Skyrmion. [Skyrme of course did pay attention to
topological and conceptual features, but apparently not in the context of
research on the collective model approach to nuclei.]
\vglue 8mm
\centerline {\large 7}
\vglue 6mm

Meghnad Saha is a major figure in the history of recent Indian science, being a
pioneer in its organization in the modern era.  He was a talented physicist
particularly well known for his work in astrophysics.  His ``Treatise on Heat''
is a splendid text book which has trained generations of Indian
students including the author.  In this short essay, an attempt has been made
to survey magnetic monopole theory with particular attention to Saha's work and
to point out that it is as important as his better known astrophysical
research.  Indeed it has had wide ranging repercussions which continue to be
influential particularly in elementary particle theory.

\vglue 8mm
\centerline{\large Acknowledgments}
\vglue 6mm
\autojoin

I am most grateful to Arshad Momen, Carl Rosenzweig, V.V. Sreedhar, Paul
Souder,  Ajit Mohan
Srivastava and Paulo Teotonio for
generous help in preparing this article.  Special thanks go to Paulo for
drawing
the figure.  This work was supported by the Department of Energy under contract
number DEFG02-85ER40231.

\vglue 8mm
\centerline{\large References}
\vglue 6mm

The references below are not meant to be exhaustive and are very limited in
number.
\begin{enumerate}
\item H. Poincar\'{e}, Compt. Rend. {\bf 123}, 530 (1896).

\item J.J. Thomson,`` Elements of the Mathematical Theory of Electricity and
Magnetism'' [Cambridge University Press, 1904].

\item A.P. Balachandran, G. Marmo, B.S. Skagerstam and A. Stern, ``Classical
Topology and Quantum States'' [World Scientific, 1991].

\item P.A.M. Dirac, {\it Proc. Roy. Soc.} (London) {\bf A133}, 60 (1931).

\item Y. Aharonov and D. Bohm, {\it Phys. Rev.} {\bf 115}, 485 (1959).

\item M.N. Saha, {\it Ind. J. Phys.} {\bf 10}, 145 (1936). Saha's derivation of
(8) was rediscovered by Wilson. See H.A. Wilson, {\it Phys. Rev.} {\bf 75}, 308
(1949) and M.N. Saha, {\it Phys. Rev.} {\bf 75}, 1968 (1949). See also M.
Fierz, {\it Helv. Phys. Acta} {\bf 17}, 27 (1944) for a closely related work.

\item A.S. Goldhaber, {\it Phys. Rev.} {\bf 36}, 1122 (1976);  R.D. Sorkin,
{\it Phys. Rev.} {\bf D27}, 1787 (1983).

\item I. Tamm, {\it Z. Physik}\, {\bf 71}, 141 (1931).

\item P. Goddard and D. Olive, {\it Rep.Prog. Phys.} {\bf 41}, 1357 (1978);  S.
Coleman, ``The Magnetic Monopole Fifty Years Later'' in ``The Unity of
Fundamental Interactions'', edited by A. Zichichi [Plenum Press, 1983].

\item V.A. Rubakov, {\it Pis'ma Zh. Eksp. Teor. Fiz.} {\bf 33}, 659 (1981)
[{\it JETP Lett.} {\bf 33}, 644 (1981)];  {\it Nucl. Phys.} {\bf B203}, 311
(1982); C.G. Callan, {\it Phys. Rev.} {\bf D25}, 2141 (1982); {\bf 26}, 2058
(1982).

\item V.G. Makhankov, Yu. P. Rybakov and V.I. Sanyuk, ``The Skyrme Model,
Fundamentals, Methods, Applications'' [Springer-Verlag (in press)].  For a
semi-popular article on the role of topology in physics, see A.P. Balachandran,
``Topology in Physics-A Perspective'', an article written in honor of Fritz
Rohrlich, Syracuse University preprint SU-4228-533 (1993), to be published in
Foundations in Physics.

\item Cf. D.J. Simms and N. Woodhouse, ``Geometric Quantization'', Lecture
Notes in Physics 53 [Springer-Verlag, 1976]; N. Woodhouse, ``Geometric
Quantization'', Oxford Mathematical Monographs [Clarendon, 1980]; J. Sniatycki,
``Geometric Quantization and Quantum Mechanics'' [Springer-Verlag, 1980].

\item Cf. L.D. Landau and M. Lifshitz, ``Quantum Mechanics, Non-Relativistic
Theory'' [Pergamon, 1977].

\item Cf. A. Bohr and B.R. Mottleson, ``Nuclear Structure, Volume II: Nuclear
Deformations'' [W.A. Benjamin, Inc., 1975]; P. Ring and P. Schuck, ``The
Nuclear Many-Body Problem'' [Springer-Verlag, 1980].

\item A.P. Balachandran, A. Simoni and D.M. Witt, {\it Int. J. Mod. Phys.}
{\bf A7}, 2087 (1992).
\end{enumerate}

\end{document}